\documentclass{ws-ijmpa}
\usepackage[super,compress]{cite}
\usepackage{hyperref}
\hypersetup{colorlinks,urlcolor=black,citecolor=black,linkcolor=black,filecolor=black}
\usepackage{breakurl}

\usepackage{graphicx}

\usepackage{tikz}
\usepackage{tikz-cd}
\usepackage{float}

\newtheorem{conj}{Conjecture}

\renewcommand*{\eqref}[1]{(\ref{#1})}

\newcommand{\arccosh}{\,\text{arccosh}\,}

\newcommand{\cicy}[2]{\begin{matrix} #1\end{matrix}\!\left[\begin{matrix}#2 \end{matrix}\right]}
\DeclareFontFamily{U}{bbold}{}
\DeclareFontShape{U}{bbold}{m}{n}
{  <-5.5> s*[1.05] bbold5
	<5.5-6.5> s*[1.05] bbold6
	<6.5-7.5> s*[1.05] bbold7
	<7.5-8.5> s*[1.05] bbold8
	<8.5-9.5> s*[1.05] bbold9
	<9.5-11.5> s*[1.05] bbold10
	<11.5-16> s*[1.05] bbold12
	<16-> s*[1.05] bbold17
}{}

\usepackage[bbgreekl]{mathbbol}
\DeclareSymbolFontAlphabet{\mathbbl}{bbold}

\newcommand{\IA}{\mathbbl{A}}

\newcommand{\IP}{\mathbbl{P}}

\newcommand{\IR}{\mathbbl{R}}

\newcommand{\IT}{\mathbbl{T}}

\newcommand{\IZ}{\mathbbl{Z}}

\newcommand{\defineas}{:=}
\newcommand{\MHV}{{\text{H}\Lambda}}
\newcommand{\HV}{{\text{HV}}}

\newcommand{\ii}{\text{i}}
\newcommand{\ee}{\text{e}}

\newcommand{\diagramscale}{0.5}

\renewcommand{\draftnote}{}

\begin{document}
\markboth{P. Kuusela and J. McGovern}{Coxeter Symmetries of GV-Invariants}

%
\catchline{}{}{}{}{}
%

\title{Reflections in the Mirror: \\[5pt] Studying Infinite Coxeter Symmetries of GV-Invariants}

\author{Pyry Kuusela}

\address{PRISMA+ Cluster of Excellence \& Mainz Institute for Theoretical Physics\\
Johannes Gutenberg-Universit\"at Mainz\\
55099 Mainz, Germany\\
pyry.r.kuusela@gmail.com}

\author{Joseph McGovern}

\address{School of Mathematics and Statistics\\
The University of Melbourne\\
Parkville, VIC 3010, Australia\\
mcgovernjv@gmail.com}

\maketitle


\begin{abstract}
We study the problem of computing Gopakumar--Vafa invariants for multiparameter families of symmetric Calabi--Yau threefolds admitting flops to diffeomorphic manifolds. There are infinite Coxeter groups, generated by permutations and flops, that act as symmetries on the GV-invariants of these manifolds. We describe how these groups are related to symmetries in GLSMs and the existence of multiple mirrors. Some representation theory of these Coxeter groups is also discussed. The symmetries provide an infinite number of relations between the GV-invariants for each fixed genus. This remarkable fact is of assistance in obtaining higher-genus invariants via the BCOV recursion. This proceedings article is based on joint work with Philip Candelas and Xenia de la~Ossa.

\keywords{Mirror symmetry; Calabi--Yau manifolds; Enumerative geometry; Multiparameter}
\end{abstract}



\section{Introduction}
We revisit the problem of finding and studying enumerative invariants of multiparameter Calabi--Yau threefolds $X$. Of particular interest to us are the \textit{Gopakumar--Vafa (GV) invariants} $n_{\mathcal{C}}^{(g)}$, which can be roughly thought of as giving the number of curves of genus $g$ in the class $\mathcal{C} \in H_2(X,\IZ)$ on $X$. These invariants play a key role, for instance, in microscopic descriptions of black holes in string theory compactifications \cite{Gaiotto:2006wm,Gaiotto:2007cd,Alexandrov:2022pgd,Alexandrov:2023zjb}. Following the seminal computation of genus-0 invariants for the quintic Calabi--Yau threefold in Ref. \citen{Candelas:1990rm}, mirror symmetry methods were further systematized and developed (see for example Ref. \citen{Cox:2000vi} for a review of these developments). The higher-genus GV-invariants can be computed via the holomorphic anomaly equations\cite{Bershadsky:1993cx,Bershadsky:1993ta}, but solving these quickly becomes computationally cumbersome for families with multiple Kähler parameters. In multiparameter cases the invariants often display a rich structure of symmetries, which can be used to significantly simplify the problem of computing the genus-$g$ GV-invariants.

Here we study a particular set of symmetries which are seen to act on the Mori cone of curves as reflections. These reflections arise, for instance, when the manifold $X$ admits a flop to a diffeomorphic manifold $X'$. Under the flop, the curves in class $\mathcal{C}$ get mapped to curves in a class $\mathcal{C}'$ which is generically different from $\mathcal{C}$. Since the GV-invariants of diffeomorphic manifolds agree when the same homology basis is used, this implies the identity $n_{\mathcal{C}}^{(g)} = n_{\mathcal{C}'}^{(g)}$. In many cases the flops can be combined with other symmetries to generate infinite groups $\mathcal{W}$, giving a highly non-trivial set of identities. By a theorem of Vinberg\cite{Vinberg_1971}, these are what are known as Coxeter~groups.

We discuss the symmetries of GV-invariants from three different, although ultimately equivalent, perspectives: In the context of gauged linear sigma models these symmetries arise when the superpotential takes on a particularly symmetric form. In purely geometric terms, the identities between invariants originate from birational transformations, which have also been studied in Refs. \citen{Brodie:2021toe,Lukas:2022crp,Hosono:2017hxe}. The symmetries can be also viewed as a consequence of the existence of multiple mirror geometries. Study of these matters dates back to Refs. \citen{Aspinwall:1993xz,Witten:1993yc,Katz:1996ht,Aspinwall:1993yb}, and has been studied in the present context for a particular family of manifolds in Ref. \citen{Hosono:2017hxe}. After this, we touch on some elements of the representation theory of Coxeter groups. As an application of these symmetries, we discuss using them to constrain higher genus holomorphic ambiguities. Computing invariants for various families with these symmetries reveals that there are repetitions of numbers that are not explained by the Coxeter symmetries, so we end with a few speculative remarks on putative larger symmetry~groups.

\section{Mirror Symmetry with Five Parameters and Coxeter Groups}	
To introduce the Coxeter symmetries, we study the GV-invariants of Kähler favorable complete intersection Calabi--Yau manifolds in products of projective spaces. Recall that the \textit{Kähler cone} of the projective space $\IP^n$ is generated by the hyperplane class $H$, so the Kähler cone $\mathcal{K}(A)$ of $A \defineas \IP^{n_0}\times \cdots \times \IP^{n_m}$ is spanned by the hyperplane classes of the individual factors, $\mathcal{K} = \langle H_0,\dots,H_m \rangle_{\IR_{\geq 0}}$. A complete intersection $X \subset A$ in the ambient space $A$ is said to be \textit{Kähler favorable}, if the Kähler cone $\mathcal{K}(X)$ of $X$ descends from that of the ambient space. Denoting by $D_i$ the restriction of $H_i$ to $X$, $\mathcal{K}(X)$ is the positive orthant in the basis given by $D_i$. We denote the curve classes dual to $D_i$ by $C^i$. These generate the \textit{Mori cone} $\mathcal{M}(X)$, the dual of the Kähler cone. Any curve class $\mathcal{C}$ can be expressed as $\mathcal{C} = p_0 C^0 + \cdots + p_m C^m$. 

Since we will always use the basis given by the $C^i$, we identify the curve class $\mathcal{C}$ with the vector $\bm{p} \defineas (p_0,\dots,p_m)$, which we call the \textit{index vector}. We call the sum of the components of $\mathbf{p}$ the (total) \textit{degree} of the curve, denoted by $\text{deg}(\mathbf{p}) \defineas i_0 + \dots + i_n$. The GV-invariant at genus-$g$ for the class $\mathcal{C}$ is denoted by $n^{(g)}_{\bm{p}}$. There is use for the notion of a \textit{positive} index vector, a vector $\mathbf{p}$ with all components non-negative and at least one component positive. Otherwise, the vector is said to be \textit{mixed}. 

For concreteness, consider the family of complete intersection Calabi--Yau manifolds\cite{Candelas:1987kf} $\MHV$ corresponding to the configuration matrix
\begin{equation}\label{eq:MHV_CICY}
\cicy{\IP^{1}\\\IP^{1}\\\IP^{1}\\\IP^{1}\\\IP^{1}}{1&1\\1&1\\1&1\\1&1\\1&1}^{h^{1,1}=5,\,h^{2,1}=45}_{\chi=-80}
\end{equation}
called \textit{mirror Hulek--Verrill manifolds}\cite{Candelas:2021lkc}. The computation of the GV-invariants of genera 0 and 1 for this family was undertaken in Ref. \citen{Candelas:2021lkc} for curve classes with total degree $< 30$. The tables of $n^{g)}_{\mathbf{p}}$ so computed show remarkable repetitions of values. 

From the CICY matrix \eqref{eq:MHV_CICY} it is immediately obvious that permuting $\IP^1$ factors of the ambient space $(\IP^1)^5$ gives a manifold belonging to the same family. This operation maps curves of genus $g$ into genus-$g$ curves, permuting the components $p_0,\dots,p_4$ of $\bm p$ so that $n^{(g)}_{\bm{p}}=n^{(g)}_{\sigma(\bm{p})}$ for any permutation $\sigma\in S_{5}$. In addition, there are identities of less obvious origin, such as $n^{(g)}_{\bm p} = n^{(g)}_{g \bm p}$ where $g$ is the map\footnote{We hope that no confusion arises between the genus-independent map $g$ and the genus of a curve.} acting~as
\begin{equation}\label{eq:MHV_inv}
g: \bm p = (p_0, \dots, p_4) \mapsto (\text{deg}(\bm p) - 2 p_0, p_1, \dots, p_4) = g \bm p~.
\end{equation}
That these identities indeed hold for all genera can be proven by noting that these arise from the fact that there exists a flop transformation mapping the manifold $\MHV$ to a diffeomorphic manifold $\MHV'$, which we will show in detail in Section \ref{sect:Birational_Transformations}.

The permutations $\sigma \in S_5$ together with the flop transformation $g$ generate a Coxeter group $\mathcal{W}$. To see this, note that the permutation groups are generated by cycles $(i,i+1)$ of length 2 and flops are always involutions. One finds a group with $n$ generators $r_{i}$, subject to relations 
\begin{equation}\label{eq:Coxeter_Presentation}
(r_{i}r_{j})^{M_{ij}}=1~,\qquad M_{ii}=1,\quad M_{ij}\in\IZ_{\geq2} \;\text{ with }i\neq j~.
\end{equation}
It follows from a theorem of Vinberg\cite{Vinberg_1971} that there are no further relations between the generators than those given above. This defines a \textit{Coxeter group} of rank $n$. These are classified by a symmetric $n\times n$ matrix $M$ of positive integers, where formal entries $M_{ij}=\infty$ indicate that the group element $r_{i}r_{j}$ is of infinite order. The information contained in the \textit{Coxeter matrix} $M$ can also be encoded in a \textit{Coxeter--Dynkin diagram}. This is a graph, consisting of one vertex for each generator $r_{i}$, an edge between all pairs of vertices $(r_{i},r_{j})$ for which $M_{ij}\geq3$, and all edges with $M_{ij}>3$ labelled by the entry $M_{ij}$. The set of all Coxeter groups contains the familiar Weyl groups from Lie theory. However, `most' Coxeter groups are infinite, and among these infinite groups are the ones of relevance to our present study. For instance, the Coxeter group associated to the family of mirror Hulek--Verrill manifolds is shown to be infinite in section \ref{sect:Orbits}.

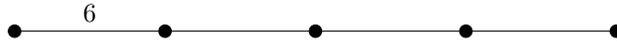
\begin{figure}[H] \label{fig:Coxeter--Dynkin_MHV}
\begin{center}
\begin{tikzpicture}[scale=2.0]

\node [style={circle, scale=\diagramscale, draw, fill=black}] (1) at ( 0.0, 0.0) {};
\node [style={circle, scale=\diagramscale, draw, fill=black}] (2) at ( 1.0, 0.0) {};
\node [style={circle, scale=\diagramscale, draw, fill=black}] (3) at ( 2.0, 0.0) {};
\node [style={circle, scale=\diagramscale, draw, fill=black}] (4) at ( 3.0, 0.0) {};
\node [style={circle, scale=\diagramscale, draw, fill=black}] (5) at ( 4.0, 0.0) {};

\draw (1) -- (2) node [midway,above] {$6$};
\draw (2) -- (3) node [midway,above] {};
\draw (3) -- (4) node [midway,above] {};
\draw (4) -- (5) node [midway,above] {};

\end{tikzpicture}
\caption{The Coxeter--Dynkin diagram giving the Coxeter symmetry group of GV-invariants of~\eqref{eq:MHV_CICY}.}
\end{center}
\end{figure}

The group under which the GV-invariants $n^{(g)}_{\bm{p}}$ of the family \eqref{eq:MHV_CICY} are invariant is generated by the cyclic permutations $(i,i+1)$, $i=1,2,3,4$, together with the involution \eqref{eq:MHV_inv}, which commutes with all of these permutations except $(1,2)$. The Coxeter diagram corresponding to these generators is displayed in Fig.~\ref{fig:Coxeter--Dynkin_MHV}.

What makes the Coxeter groups generated by flops and permutations so powerful for analyzing the enumerative invariants is that in many cases these groups are infinite, thus providing infinitely many equalities between the invariants. 

\section{Coxeter Symmetries from GLSMs, Flops, and Mirror Symmetry}\label{sect:Birational_Transformations}
The existence of the Coxeter symmetries that we study requires an identity analogous to \eqref{eq:MHV_inv} which, together with the permutation (or other similar) identities, gives rise to an infinite group. Such identities arise when a Calabi--Yau manifold has a flop to a diffeomorphic manifold \cite{Brodie:2021toe}. This can be studied from several perspectives, three of which we will briefly review in this section.

Consider again, for concreteness, the example of the mirror Hulek--Verrill manifolds --- the generalisation to other cases should be clear. These are defined as the vanishing loci of two multilinear polynomials in $(\IP^1)^5$:
\begin{align} \label{eq:MHV_polynomials}
G_{a_0}(X) = \sum_{\substack{a_j = 0\\j \neq 0}}^{1} A_{a_0,a_1,a_2,a_3,a_4,a_5} X_{1,a_1} X_{2,a_2} X_{3,a_3}X_{4,a_4} X_{5,a_5}~, \qquad a_0=0,1~,
\end{align}
where we have introduced the homogeneous coordinates $[X_{j,0}:X_{j,1}]$ on the $j$'th copy of $\IP^1$. It will be useful to write these two equations in matrix form as
\begin{align} \nonumber
&\left( \begin{matrix}
\;\IA_{0,0} \;\; & \IA_{0,1} \; \\
\;\IA_{1,0} \;\;& \IA_{1,1} \;
\end{matrix}  \right) \left( \begin{matrix}
X_{1,0}\\
X_{1,1}
\end{matrix} \right) \defineas A X_1 = 0~, \qquad \text{where}\\
&\hskip25pt \IA_{b,c} \defineas \sum_{a_j = 0}^{1} A_{b,c,a_2,a_3,a_4,a_5} X_{2,a_2} X_{3,a_3}X_{4,a_4} X_{5,a_5}~, \label{eq:MHV_matrix_equations}
\end{align}
and the second equation on the first line defines the matrix $A$ and the vector $X_1$. Here we have singled out the coordinate $X_1$, although we could have chosen any $X_i$ to obtain flops along other copies of $\IP^1$.
\subsection{Coxeter Symmetries from GLSMs}
The structures that lead to the Coxeter symmetries between the GV-invariants can be exhibited in a GLSM setup with the gauge group $G = U(1)^5$, fields $X_{\mu,a}$ with $0\leq\mu\leq5$ and $0\leq a\leq1$, and charge assignments and Fayet–-Iliopoulos-parameters as given in Table \ref{tab:GLSM_Matrix_1}. In addition, the $U(1)_R$ charges are assigned so that the fields $X_{0,a}$ have charge $-2$ while the other fields $X_{i,a}$ have charge $0$. The superpotential takes a particularly symmetric form, being given by 
\begin{align}\label{eq:superpot}
W {=} \sum_{a_j = 0}^{1} A_{a_0,\dots,a_5} X_{0,a_0}X_{1,a_1}\cdots X_{5,a_5}=X_{0,0} G_0(X) {+} X_{0,1} G_1(X) = X_0^T A X_1~.
\end{align}
To obtain the third equality we have introduced a vector $X_0=(X_{0,0},X_{0,1})^T$.  It would be interesting to investigate whether this symmetry has interesting consequences for the UV theory. Our present interest lies in symmetries of the instanton numbers, so it is sufficient to study the IR theory.
\begin{table}[H]
\tbl{\label{tab:GLSM_Matrix_1}
The $U(1)$ charges and FI-terms of the GLSM that realizes the mirror Hulek--Verrill manifold \eqref{eq:MHV_polynomials} in the phase where all FI-parameters are positive.}
{
\begin{tabular}
{c||p{0.4cm}p{0.4cm}|p{0.4cm}p{0.4cm}|p{0.4cm}p{0.4cm}|p{0.4cm}p{0.4cm}|p{0.4cm}p{0.4cm}||p{0.4cm}p{0.4cm}||c}
  & \!\!$X_{1,0}$ & \!\!$X_{1,1}$ & \!\!$X_{2,0}$ & \!\!$X_{2,1}$ & \!\!$X_{3,0}$ & \!\!$X_{3,1}$  &\!\!$X_{4,0}$ & \!\!$X_{4,1}$  & \!\!$X_{5,0}$ & \!\!$X_{5,1}$ & $X_{0,0}$ & $X_{0,1}$ & FI\\ \hline
\!$U(1)_1$ & $1$ & $1$ & $0$ & $0$ &  $0$ & $0$ & $0$ & $0$ & $0$ & $0$ & $-1$ & $-1$ & $\xi_1$\\
\!$U(1)_2$ & $0$ & $0$ & $1$ & $1$ &  $0$ & $0$ & $0$ & $0$ & $0$ & $0$ & $-1$ & $-1$ & $\xi_2$\\
\!$U(1)_3$ & $0$ & $0$ & $0$ & $0$ &  $1$ & $1$ & $0$ & $0$ & $0$ & $0$ & $-1$ & $-1$ & $\xi_3$ \\
\!$U(1)_4$ & $0$ & $0$ & $0$ & $0$ &  $0$ & $0$ & $1$ & $1$ & $0$ & $0$ & $-1$ & $-1$ & $\xi_4$ \\
\!$U(1)_5$ & $0$ & $0$ & $0$ & $0$ &  $0$ & $0$ & $0$ & $0$ & $1$ & $1$ & $-1$ & $-1$ & $\xi_5$ \\
\end{tabular}
}
\end{table}
\noindent The D-term equations are
\begin{align}
|X_{i,0}|^2 + |X_{i,1}|^2 - |X_{0,0}|^2 - |X_{0,1}|^2 = \xi_i~. 
\end{align}
In the phase $\xi_i > 0$ this model realizes the mirror Hulek--Verrill manifold defined by the vanishing of the polynomials \eqref{eq:MHV_polynomials}, with the fields $X_{i,0}$, $X_{i,1}$ corresponding to the homogenous coordinates of the ambient space $(\IP^1)^5$, and the fields $X_{0,0},X_{0,1}$ corresponding to a bundle $\mathcal{O}(-1,-1,-1,-1,-1) \oplus \mathcal{O}(-1,-1,-1,-1,-1)$ over $(\IP^1)^5$.

Consider then the phase with $\xi_1 < 0$ and $\xi_i >0$ for $1<i\leq5$. In this phase, taking suitable linear combinations of the $D$-term equations gives
\begin{align}
|X_{0,0}|^2 + |X_{0,1}|^2 - |X_{1,0}|^2 - |X_{1,1}|^2 = \phantom{\xi_i}-\xi_1 \defineas \xi_1' >0~,\\
|X_{i,0}|^2 + |X_{i,1}|^2 - |X_{1,0}|^2 - |X_{1,1}|^2 = \xi_i-\xi_1 \defineas \xi_i' >0~.
\end{align}
By redefining the charges similarly, we end up with the data in Table \ref{tab:GLSM_Matrix_2}. Additionally, by noting that the $U(1)_R$ charge of a field is defined up to addition of the charge under the global $U(1)^5$ symmetry, we can redefine the $U(1)_R$ charges so that now $X_{1,a}$ has charge $-2$ while the other fields  $X_{\mu,a}$, $\mu \neq 1$, have charge $0$.
\begin{table}[H]
\tbl{\label{tab:GLSM_Matrix_2}
The $U(1)$ charges and FI-parameters after taking suitable linear combinations, so that  $\xi_i'>0$, and with the charges defined so that they make the geometric interpretation apparent. Note the same structure as Table \ref{tab:GLSM_Matrix_1}, but with $X_{0,a}$ and $X_{1,a}$ exchanged.}
{
\begin{tabular}{c||p{0.4cm}p{0.4cm}|p{0.4cm}p{0.4cm}|p{0.4cm}p{0.4cm}|p{0.4cm}p{0.4cm}|p{0.4cm}p{0.4cm}||p{0.4cm}p{0.4cm}||c}
  & \!\!$X_{1,0}$ & \!\!$X_{1,1}$ & \!\!$X_{2,0}$ & \!\!$X_{2,1}$ & \!\!$X_{3,0}$ & \!\!$X_{3,1}$  &\!\!$X_{4,0}$ & \!\!$X_{4,1}$  & \!\!$X_{5,0}$ & \!\!$X_{5,1}$ & $X_{0,0}$ & $X_{0,1}$ & FI\\ \hline
\!$U(1)_1$ & $-1$ & $-1$ & $0$ & $0$ &  $0$ & $0$ & $0$ & $0$ & $0$ & $0$ & $1$ & $1$ & $\xi_1'$\\
\!$U(1)_2$ & $-1$ & $-1$ & $1$ & $1$ &  $0$ & $0$ & $0$ & $0$ & $0$ & $0$ & $0$ & $0$ & $\xi_2'$\\
\!$U(1)_3$ & $-1$ & $-1$ & $0$ & $0$ &  $1$ & $1$ & $0$ & $0$ & $0$ & $0$ & $0$ & $0$ & $\xi_3'$ \\
\!$U(1)_4$ & $-1$ & $-1$ & $0$ & $0$ &  $0$ & $0$ & $1$ & $1$ & $0$ & $0$ & $0$ & $0$ & $\xi_4'$ \\
\!$U(1)_5$ & $-1$ & $-1$ & $0$ & $0$ &  $0$ & $0$ & $0$ & $0$ & $1$ & $1$ & $0$ & $0$ & $\xi_5'$ \\
\end{tabular}
}
\end{table}
We see that the fields $X_{1,a}$ and $X_{0,a}$ have changed roles. Geometrically the theory corresponds again to a complete intersection in $(\IP^1)^5$ but with the homogeneous coordinates of the ambient space being $(X_{\mu,0},X_{\mu,1})$ with $\mu \neq 1$. Now $X_{1,0}$ and $X_{1,1}$ give the coordinates on the bundle ${\mathcal{O}(-1,-1,-1,-1,-1) \oplus \mathcal{O}(-1,-1,-1,-1,-1)}$. The equations defining the manifold are given by
\begin{align} \label{eq:MHV_polynomials2}
G_{a_1}'(X) = \sum_{\substack{a_j = 0\\j \neq 1}}^{1} A_{a_0,a_1,a_2,a_3,a_4,a_5} X_{0,a_0} X_{2,a_2} X_{3,a_3}X_{4,a_4} X_{5,a_5}~, \qquad a_1=0,1~,
\end{align}
or in matrix form, simply as
\begin{align}
A^{\,T} X_0 = 0~.
\end{align}
Crucially, the complete intersection $\MHV'$ defined by the condition that the polynomials $G_0'(X)$ and $G_1'(X)$ vanish belongs to the family of CICYs corresponding to the matrix \eqref{eq:MHV_CICY}. That is, $\MHV'$ is diffeomorphic to $\MHV$. This is required to obtain relations between different GV-invariants of a single family of manifolds.

\subsection{Coxeter Symmetries from Flop Transitions} \label{sect:flops}
\begin{figure}[H]
\centering
\begin{tikzcd}
\MHV \arrow[rr, leftrightarrow, dashed] \arrow[dr,"\pi_1" ']  &   & \MHV' \arrow[rr, leftrightarrow, dashed]  \arrow[dr,"\pi_2" ']& & \MHV'' \arrow[rr, leftarrow, dashed]  \arrow[dr,"\pi_2" '] & & \cdots\\
         				  & \widehat{\MHV} \arrow[ur, leftarrow,"\pi_0" '] & & \widehat{\MHV}' \arrow[ur, leftarrow,"\pi_3" '] & & \cdots\\
\end{tikzcd}
\caption{An infinite chain of flops from $\MHV$ to diffeomorphic threefolds $\MHV',\MHV'',$ etc. The first flop is obtained by using the projection $\pi_1$ along the first copy of $\IP^1$ in the ambient space $(\IP^1)^5 \supset \MHV$ to a singular manifold $\widehat{\MHV}$. The same singular manifold can be obtained by using the projection $\pi_0$ along the first $\IP^1$ in the ambient space of another threefold $\MHV' \subset (\IP^1)^5$ diffeomorphic to $\MHV$. In this way one obtains a birational map between $\MHV$ and $\MHV'$ that flops the curves in the class $\bm{e}_1 = (1,0,0,0,0)$. Another flop is obtained by using projection $\pi_2$ along the second $\IP^1$ of the ambient space of $\MHV'$. This process can be continued to obtain an infinite number of distinct flops. A similar chain of flops for the family of split quintic Calabi--Yau threefolds has been studied in detail in Ref. \citen{Hosono:2017hxe}.\label{fig:flops}}
\end{figure}
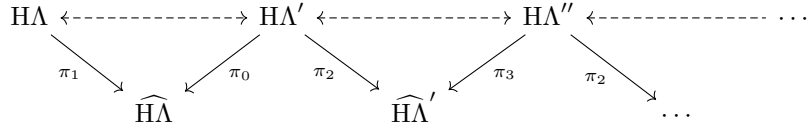

In purely geometric terms the above procedure gives a flop transition of $\MHV$ to a birational diffeomorphic manifold $\MHV'$. To see how this comes about, note that from the matrix equation \eqref{eq:MHV_matrix_equations}, it follows that the manifold $\MHV$ is birational to the \textit{contraction}\cite{Candelas:1987kf} of $\MHV$, $\widehat{\MHV}$, which is defined by the condition $\text{det}(A) = 0$. The birational map is given by the projection $\pi_1$ along the first $\IP^1$ in the ambient product space with coordinates $X_{1,0}$ and $X_{1,1}$.
The same argument shows that the manifold $\MHV'$ which corresponds to the transpose of $A$, being defined by $A^T X_0 = 0$, is birational to $\widehat{\MHV}$, and thus to $\MHV$. We denote the projection $\MHV \to \widehat{\MHV}$ along the first $\IP^1$ by $\pi_0$. As mentioned above, $\MHV'$ is diffeomorphic to $\MHV$, and by carefully studying the transformation, one can show that this is in fact a flop\cite{Brodie:2021toe}. By flopping curves in different classes, or equivalently, by using this flop together with the $S_5$ permutation symmetries, one can obtain an infinite sequence of flops (see Fig. \ref{fig:flops}).
To see that this transformation is responsible for the identities \eqref{eq:MHV_inv}, one can study the action of the birational map $\MHV \to \MHV'$ on curves explicitly: Consider a curve $\gamma \subset \HV$, of genus zero and degree $\bm p=(p_0,\dots,p_4)$. The curve $\gamma$ projects under $\pi_1$ to a curve $\pi_1(\gamma) \defineas \widehat{\gamma}$ of degree $\widehat{\bm{p}} = (p_1,\dots,p_4)$, unless $\gamma$ is one of the curves contracted under the flop transformation, which in this case are those in the curve classes $n C^0$, $n \in \IZ$ lying on the boundary of the Mori cone. The curve $\widehat{\gamma}$ lifts to a curve $\gamma'$ of degree $\bm{p}'$ in $\MHV'$ under $\pi_0^{-1}$. This gives a bijective map between the curves of $\MHV$ and $\MHV'$, which implies that the GV-invariants $n^{(0)}_{\bm{p}}$ and $n^{(0)}_{\bm{p}'}$ must agree. We will now show that this relation is exactly \eqref{eq:MHV_inv}. 

Consider $\gamma$ given by an embedding $\IP^1 \hookrightarrow \MHV \subset (\IP^1)^5$ with components
\begin{align} \label{eq:embidding}
(X_{1,1}(z)/X_{1,0}(z),\dots,X_{5,1}(z)/X_{5,0}(z))~,
\end{align}
where $z$ denotes the affine coordinate on the curve $\gamma$, and we use the affine coordinates for $(\IP^1)^5$. Thus, on $\widehat \gamma$, we may take the coordinates $X_{j,a}$, $j=2,\dots,5$, $a=0,1$ to be polynomials in $z$ which are of minimal degree in the sense that $X_{j,1}(z)$ and $X_{j,0}(z)$ have no common factors. 

By looking at the equations \eqref{eq:MHV_matrix_equations} defining the manifold $\MHV$, we have that the first component in \eqref{eq:embidding} of the lift $\gamma = \pi_1^{-1}(\widehat{\gamma})$ is given by
\begin{align}
\label{eq:ratios}
\frac{X_{1,1}(z)}{X_{1,0}(z)} = -\frac{\IA_{00}(z)}{\IA_{01}(z)} \;= -\frac{\IA_{10}(z)}{\IA_{11}(z)}~.
\end{align}
Similarly, the first component of the lift $\gamma' = \pi_0^{-1}(\widehat{\gamma})$ is given by
\begin{align}
\frac{X_{0,1}(z)}{X_{0,0}(z)} = -\frac{\IA_{00}(z)}{\IA_{10}(z)} \;= -\frac{\IA_{01}(z)}{\IA_{11}(z)} ~.
\end{align}
By assumption $\gamma$ is in curve class $\bm{p}$, so the ratio $X_{1,1}(z)/X_{1,0}(z)$ must be of degree $p_0$, implying that $\IA_{00}(z)$ and $\IA_{01}(z)$ must have $p_1+p_2+p_3+p_4-p_0$ factors in common. A simple counting argument shows \cite{Candelas:2021lkc} that $\IA_{00}(z)$ and $\IA_{10}(z)$ must then have $p_0$ factors in common, implying that $\gamma'$ is in the curve class $g \bm{p}$.

Since the action of the flop on the curve classes is linear, the Mori cone of curves on $\MHV$, $\mathcal{M}(\MHV)$ also transforms under the action of the flop. Explicitly, 
\begin{align}
\mathcal{M}(\MHV) = \{a_0 C^0 + \cdots a_4 C^4 \; | \; a_i \geq 0 \} \simeq \{(a_0,\dots, a_4) \; | \; a_i \geq 0\}
\end{align}
gets mapped by $g$ to
\begin{align} \nonumber
g\mathcal{M}(\MHV) &\simeq \mathcal{M}(\MHV') = \{a_0 C_0' + \cdots + a_4 C_4' \; | \; a_i \geq 0 \}\\
&\simeq \{(-a_0+a_1+\cdots + a_4, a_1, \dots ,a_4) \; | \; a_i \geq 0\}~,
\end{align}
where the second equality follows from the fact that $C_0' = -C_0$ and $C_i' = C_0 + C_i$ for $i \neq 0$. If a curve is not flopped, and it maps to outside of the Mori cone, its GV-invariant must vanish. This motivates defining what was called in Ref. \citen{Lukas:2022crp} the \textit{restricted Mori cone} as the intersection of all images of the Mori cone $\mathcal{M}(\MHV)$ under the action of the Coxeter group $\mathcal{W}$:
\begin{align}
\mathcal{M}_{\text{restr}}(\MHV) \defineas \bigcap_{w \in \mathcal{W}} w\mathcal{M}(\MHV)~.
\end{align}
Only the curve classes in the restricted cone, or $\bm{e}_1$ and its positive images, can have non-vanishing GV-invariant.

\begin{figure}[H]
	\centering
	\begin{tikzpicture}[scale=0.85, every node/.style={scale=1}]
	\usetikzlibrary{arrows.meta}
	
	\fill[gray!30] (0,0) -- (3,3) -- (0,3) -- cycle;

    \draw[line width=0.4mm,gray,-stealth] (0,-0.5)--(0,3);
    \draw[line width=0.4mm,gray,-stealth] (-3,0)--(3,0);

    \draw[line width=0.4mm,black,-stealth] (0,0)--(0,1.5);
    \draw[line width=0.4mm,black,-stealth] (0,0)--(1.5,0);  

    \draw[line width=0.4mm,black,-stealth] (0,0)--(1.5,1.5);  
    \draw[line width=0.4mm,black,-stealth] (0,0)--(-1.5,0);    	
	
	\foreach \Point in {(0.75,0),(1.5,0)}{
	   \node at \Point {\textbullet};
	}

 	\foreach \Point in {(2.25,0),(3,0),(1.5,0.75),(2.25,0.75),(3,0.75),(2.25,1.5),(3,1.5),(3,2.25)}{
	   \node at \Point {$\bm{\times}$};
	}

	\end{tikzpicture}
	\caption{A cartoon of the Mori and restricted cones in a two-parameter analogue of the above discussion with only one flop corresponding to an involution ${\bm p = (p_0,p_1) \mapsto g \bm p = (p_1-p_0,p_1)}$. The black arrows represent the generators of the Mori cones ${\mathcal{M} = \langle (0,1), (1,0) \rangle_{\IR_{\geq 0}}}$ and ${g\mathcal{M} = \langle (-1,0), (1,1) \rangle_{\IR_{\geq 0}}}$. The restricted cone $\mathcal{M}_{\text{restr}} = \mathcal{M} \cap g\mathcal{M}$ is represented by the shaded region. The dots represent curve classes that are flopped. It can be shown \cite{} that in such classes only $(1,0)$ and $(2,0)$ have non-zero GV-invariants. The crosses represent curve classes that belong to the Mori cone $\mathcal{M}$ but lie outside of the restricted cone, and therefore have vanishing GV-invariants.}	
\end{figure}
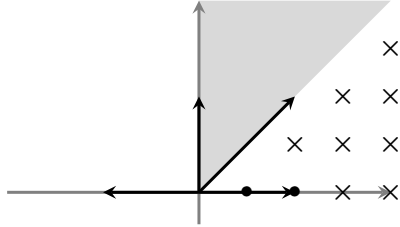
Analogous considerations apply also to the dual divisor classes $D_i$ and the corresponding Kähler cone. From the action of the flop on the divisors, one can derive the action of the Coxeter group on the Picard group and thus on the Kähler cone (for details, see Ref. \citen{Lukas:2022crp}), which is simply the dual to the action on the curve classes. This explains why the Coxeter symmetry applies to the higher-genus invariants as well. One can see these relations as arising from keeping the curves fixed and flopping the divisors, so the curve classes must transform in the same way for all genera. 

The dual to the restricted cone is the \textit{extended Kähler cone}, the union of all images of the Kähler cone $\mathcal{K}(\MHV)$ under the action of the Coxeter group:
\begin{align}
\mathcal{K}_{\text{ext}}(\MHV) \defineas \bigcup_{w \in \mathcal{W}} w\mathcal{K}(\MHV)~.
\end{align}

\subsection{Coxeter Symmetries from Mirror Symmetry} \label{sect:mirror_symmetry}
Symmetries of the GV-invariants can be also studied from the perspective of mirror symmetry. The family $\MHV$ are mirrors to the family of Hulek--Verill threefolds\footnote{In this section, we sometimes display the moduli of the manifolds $\HV$ explicitly to avoid confusion.} $\HV_{(\varphi^0,\dots,\varphi^5)}$ with five complex structure parameters\cite{Hulek2005}, defined as toric compactifications of the varieties given by the vanishing of the two Laurent polynomials
\begin{align}\label{eq:TwoPolynomials} 
P^1(\mathbf{Y}) = \sum_{\mu=0}^5 Y_\mu~, \hskip30pt P^2(\mathbf{Y};\varphi) = \sum_{\mu=0}^5 \frac{\varphi^\mu}{X_\mu}~,
\end{align}
on the projective torus $\IT^5 \defineas \IP^5/\{Y_0\cdots Y_5 = 0\}$. The $\varphi^\mu$, $\mu=0,1,\dots,5$ furnish projective coordinates of the complex structure moduli space of $\HV$ that enter on an equal footing. This implies in particular that two manifolds whose complex structure parameters $[\varphi^0:\dots:\varphi^5]$ are related by an $S_6$ permutation are biholomorphic. This family of manifolds has six large complex structure singularities at the intersections $E_0 \cap E_1 \cap \cdots \cap \widehat{E_\mu} \cap \cdots \cap E_5$ of the divisors $E_\nu := \{\varphi^\nu = 0 \}$, where $E_\mu$ is omitted.

In this case mirror symmetry associates to a manifold $\HV$ near a large complex structure point $\varphi_*$ not a single manifold $\MHV$, but rather the set of manifolds that are related to $\MHV$ by a flop of the type described above. One sees this explicitly by studying the mirror map: Choose an integral basis $(A^a,B_a)$ of the middle homology $H_3(\HV,\IZ)$ with a canonical dual basis $(\alpha_a,\beta^a)$, $a=0,\dots,h^{1,2}$, that satisfy
\begin{align} \label{eq:symplectic_basis}
\int_{A^b} \alpha_a = \int_X \alpha_a \wedge \beta^b = \delta_a^b~, \qquad \int_{B_a} \beta^b = \int_X \beta^b \wedge \alpha_a = -\delta_a^b~.
\end{align}
\textit{Periods} of $X$ are integrals of the holomorphic $(3,0)$-form $\Omega$ over the cycles $(A^a,B_a)$:
\begin{align}
z^a \defineas \int_{A^a} \Omega~, \qquad \mathcal{G}_a \defineas \int_{B_a} \Omega~.
\end{align}
It can be shown that there exists a \textit{prepotential} $\mathcal{G}(z)$,  a homogeneous function of degree 2 in $z^a$  so that $\mathcal{G}_a = \partial_{z^a} \mathcal{G}(z)$. Using topological quantities of the mirror family $\MHV$, one can also define a function $\mathcal{F}$ of the complexified Kähler parameters $t^i$ of~$\MHV$:
\begin{equation} \label{eq:HV_prepotential}
\mathcal{F}(\bm t)\defineas \frac{1}{6}Y_{ijk}\,t^{i}t^{j}t^{k}-\frac{1}{24}c_{2,i}\,t^{i}-\frac{\zeta(3)}{2(2\pi\text{i})^{3}}\,\chi +\frac{1}{(2\pi \text{i})^{3}}\sum_{\bm{p}}n_{\bm{p}} \, \text{Li}_{3}\left(q^{\bm{p}}\right)~,  
\end{equation}
where $Y_{ijk}$ are the triple intersection numbers, $c_{2,i}$ the second Chern numbers, and ${q^{\bm{p}} \defineas \ee^{2\pi \ii \bm{t} \cdot \bm{p}}}$. The sum is over \textit{positive} index vectors $\bm{p}$.\footnote{An index vector $\bm{p}$ is positive if all of its components are non-negative, with at least one non-zero.} 

The \textit{mirror map} is given by identifying the \textit{flat coordinates} $z^i/z^0$ with the complexified Kähler parameters $t^i$ of the family of mirror manifolds $\MHV$. Under this identification, one can also identify $\mathcal{F}$ with the prepotential $\mathcal{G}$, 
\begin{align}
\frac{\mathcal{G}(z)}{-\big(z^0\big)^2} \simeq \mathcal{F}(\bm t)
\end{align}
where $\simeq$ denotes that identification holds up to an effect of symplectic transformations of the basis \eqref{eq:symplectic_basis}. These can be seen to act on the polynomial part of $\mathcal{F}$, and arise from the fact that the symplectic integral basis \eqref{eq:symplectic_basis} is not uniquely determined. 

Let $\bm 0$ denote the large complex structure point at ${[1:0:0:0:0:0]}$, and $\widetilde{\bm 0}$ denote the large complex structure point at ${[0:1:0:0:0:0]}$. Analytic continuation\cite{Candelas:2021lkc} relates the flat coordinates $z^i/z^0$ near $\bm 0$ to the flat coordinates $\widetilde{z}^{1}/\widetilde{z}^{0}$:
\begin{equation} \label{eq:changes}
\widetilde{z}^{1}/\widetilde{z}^{0} =-z^{1}/z^0~,\qquad \widetilde{z}^i/\widetilde{z}^0 =z^{i}/z^0+z^{1}/z^0~,\quad i>1~.
\end{equation}
This gives a relation between the mirror $\MHV_{\bm t}$ of a manifold $\HV_{(1,\varphi^1,\dots,\varphi^5)}$ near $\bm 0$ and the mirror $\MHV_{\widetilde{\bm{t}}}$ of $\HV_{(\varphi^1,1,\varphi^2,\dots,\varphi^5)}$ near the other large complex structure point $\widetilde{\bm 0}$. Namely, $\MHV_{\bm t}$ and $\MHV_{\widetilde{\bm t}}$ are related by the flop discussed above. However, as the manifolds $\HV_{(1,\varphi^1,\dots,\varphi^5)}$ and $\HV_{(\varphi^1,1,\varphi^2,\dots,\varphi^5)}$ are biholomorphic, from the perspective of mirror symmetry they can be identified. Therefore we can view both $\MHV_{\bm t}$ and $\MHV_{\widetilde{\bm t}}$ as mirrors of $\HV_{(1,\varphi^1,\dots,\varphi^5)}$ but each with a different mirror map, which are related by \eqref{eq:changes}. This means we should be able to write $\mathcal{G}$ in terms of both $\mathcal{F}$ and $\widetilde{\mathcal{F}}$ associated to $\MHV_{\widetilde{\bm t}}$: 
\begin{align}\label{eq:prepotential_relation}
\mathcal{G}(z) \simeq \mathcal{F}(\bm t) \simeq \widetilde{\mathcal{F}}(\widetilde{\bm t})~.
\end{align}
As the topological quantities appearing in \eqref{eq:HV_prepotential} are invariant under the flop, we have in fact that the functional forms of $\widetilde{\mathcal{F}}$ and $\mathcal{F}$ are identical. Hence the relation \eqref{eq:prepotential_relation} implies that $\mathcal{F}(\bm t)$ is equal to $\mathcal{F}(\widetilde{\bm t})$, up to the action of a symplectic transformation.

The relations $\widetilde{t}^1 = - t^1$, $\widetilde{t}^i = t^i + t^1$, $i=1,\dots,4$, following from \eqref{eq:changes}, give
\begin{equation}\label{eq:prepot2}
\mathcal{F}(\bm t) = \mathcal{F}\big( \,\bm{\widetilde{t}} \; \big) - 4 \left(\, \widetilde{t^1} \, \right)^3 - 2 \widetilde{t^1}~.
\end{equation}
Requiring that the both sides of the equation are equal up to the effect of a symplectic transformation gives the condition
\begin{equation}\label{eq:uptosymplectic1}
4t_{1}^{3}+2t_{1}-\frac{1}{(2\pi\text{i})^{3}}\sum_{\bm p}n_{\bm p}\text{Li}_{3}\!\left(q^{\widetilde{\bm p}}\right) \simeq -\frac{1}{(2\pi\text{i})^{3}}\sum_{\bm p}n_{\bm p}\text{Li}_{3}\!\left(q^{\bm p}\right)~.
\end{equation}
Breaking up the sum over $\bm p$ into a sum over index vectors $\bm p$ for which both $\bm p$ and $\widetilde{\bm p}$ are positive and $\bm{e}_1 = (1,0,0,0,0)$ (for which $\widetilde{\bm e}_1 = (-1,0,0,0,0)$), one obtains
\begin{equation} \label{eq:uptosymplectic2}
4t_{1}^{3}+2t_{1} \simeq \frac{1}{(2\pi\ii)^{3}}n_{\bm e_1}\left(\text{Li}_{3}\!\left(q_{1}^{-1}\right)-\text{Li}_{3}\!\left(q_{1}\right)\right)-\frac{1}{(2\pi\ii)^{3}}\sum_{\substack{\bm p \text{ for which}\\ \bm p,\,\widetilde{\bm p}>0}}\!\!\!\!\!(n_{\bm p}-n_{\widetilde{\bm p}})\text{Li}_{3}\!\left(\bm q^{\bm p}\right)~.
\end{equation}
The symplectic transformations can only affect the polynomial terms, so the equality up to a symplectic transformation imposes that $n_{\bm p}=n_{\widetilde{\bm p}}$ for all positive $\bm p$ such that $\widetilde{\bm p}$ is also positive. This is nothing but the Coxeter symmetry relation \eqref{eq:MHV_inv}. In addition, this requires that $n_{\bm{e}_1}=24$, which follows from the identity
\begin{equation} \notag
\frac{24}{(2\pi \ii)^{3}}\bigg(\text{Li}_{3}\!\left(\ee^{-2\pi \ii t_{1}}\right)-\text{Li}_{3}\!\left(\ee^{2\pi\ii t_{1}}\right)\bigg) = 2t_{1}-6t_{1}^{2}+4t_{1}^{3}~,\qquad \text{Im}[t_{1}]>0~.
\end{equation}
The term $-6t_{1}^{2}$ not appearing in \eqref{eq:uptosymplectic2} is removed by a change of symplectic basis of $H^3(X,\IZ)$. Similar arguments apply, mutatis mutandis, to higher-genus invariants.

\subsection{Existence of Flops and Coxeter-Symmetric CICYs}
For complete intersection Calabi--Yau manifolds there is a simple criterion that can be used to find families where these flops to diffeomorphic manifolds exist. Such flops were classified for Kähler favorable CICY threefolds in Ref. \citen{Brodie:2021toe}, dividing them into two cases based on the form of the configuration matrix:
\begin{align}
&\textbf{Type I}: & &\textbf{Type II}: \nonumber \\
&\cicy{\IP^{n}\\\overline{\IP}}{1& 1 & \cdots & 1 & 0 & \cdots & 0\\\bm{q} & \bm{q} & \cdots & \bm{q} & \bm{q}_1 & \cdots  &\bm{q}_k }& & 
\cicy{\IP^{n}\\\overline{\IP}}{2 & 1 & \cdots & 1 & 0 & \cdots & 0\\\bm{q}_1 & \bm{q}_2 & \cdots & \bm{q}_n & \bm{q}_{n+1} & \cdots  &\bm{q}_k }
\end{align}
where $\overline{\IP}$ denotes the product space $\IP^{n_1} \times \cdots \times \IP^{n_m}$ and the $\bm{q}_i$ are $m$-component column vectors. 

Type I corresponds to the cases discussed previously, and the results in section \ref{sect:flops} apply with obvious modifications. The action on the curve classes $\bm p$ is given\cite{Lukas:2022crp}~by
\begin{align}
\bm{p} = (p_0, \dots, p_m) \mapsto \left( n \bm{q} \cdot \widehat{\bm{p}} - p_0, p_1, \dots, p_m \right)~,
\end{align}
where, as above, $\widehat{\bm p} = (p_1,\dots,p_m)$. For type II flops the action is given\cite{Lukas:2022crp} by
\begin{align}
\bm{p} = (p_0, \dots, p_m) \mapsto \left(\left(\bm{q}_1+2\textstyle \sum_{i=2}^{n} \bm{q}_i\right) \cdot \widehat{\bm{p}}-p_0, p_1,\dots,p_m \right)~.
\end{align}

The existence of flops of type I and II are not mutually exclusive. For instance, 
\begin{equation}\label{eq:special}
\cicy{\IP^{3}\\\IP^{5}}{0&1&1&1&1\\2&1&1&1&1}
\end{equation}
has two rows, one of type I and one of type II. Indeed, this is another way in which one can obtain infinite Coxeter groups if the two flops do not commute with each other. For the family \eqref{eq:special}, the two flops generate the infinite dihedral group.

These two types of flops, together with possible permutation symmetries, already give a large set of Coxeter groups. However, these symmetries are not restricted to flops nor to complete intersections in projective spaces. For instance, there are symmetries arising from birational transformations of  complete intersections in weighted projective spaces. More speculatively, the conjectural GV-invariants in the non-toric ``Phase IV" geometry of the non-Abelian GLSM studied in Ref. \citen{Hori:2016txh} display a $\IZ_2$ symmetry that has properties expected of a symmetry arising from a flop: the invariants are symmetric under $(p_0,p_1) \mapsto (p_0,5p_0-p_1)$, at least up to total degree 31 at genera 0 and 1, and the only `flopping' classes $(0,p)$ with non-vanishing genus-0 invariants are $(0,1)$ and $(0,2)$. While the invariants so computed for this non-Abelian model remain strictly speaking conjectural, we anticipate that geometrical considerations related to Coxeter symmetries will place the numbers of Ref. \citen{Hori:2016txh} on firmer footing.

\section{Coxeter Group Actions and Their Orbits} \label{sect:Orbits}
Certain aspects of the representations of the Coxeter groups are important for studying structures arising in the set of invariants. For instance, the structure of the restricted cone, the extended Kähler cone, and the modularity properties of the prepotential $\mathcal{F}$ in the Coxeter-symmetric cases\cite{Lukas:2022crp} are all related to the structure of these representations. Most of this discussion can be framed in an elementary manner. However, seemingly simple statements turn out to be difficult to prove. 

We consider the representations of Coxeter groups $\mathcal{W}$ that are given by the permutations of the components of $\mathbf{p}$ together with the `flop' action
\begin{align}
g: \mathbf{p} = (p_0,\dots,p_n) \mapsto (k(p_1+\dots+p_n) - p_0, p_1, \dots, p_n)~.
\end{align}
As can be seen from the discussion above, this is not the most general form of the action that a Coxeter group generated by flops and permutation symmetries can have. However, this form already accounts for a large number of examples, and allows us to make some concrete statements that would become much more cumbersome to formulate in the most general case.

Let us introduce some notation. Given any integral vector $\mathbf{p}$, we have the orbit
\begin{equation}
\mathcal{W}[\mathbf{p}]=\{w(\mathbf{p})\;|\; w\in\mathcal{W}\}~.
\end{equation}
We call an orbit $\mathcal{W}[\mathbf{p}]$ positive/mixed if every vector in the orbit is positive/mixed. 

By definition, the GV-invariants corresponding to two index vectors in the same orbit agree. In particular, $n_{\bm{p}}$ can be non-vanishing only if $\mathbf{p}$ belongs to a positive orbit or to the `half-orbit' $\mathcal{W}^+ \defineas \{w \bm{e}_1 >0 \; | \; w\in\mathcal{W} \}$ of the flopping curve classes. The latter can have non-zero GV-invariant, even though the mixed vectors in the orbit have vanishing invariants. In fact, this is the statement that the positive orbits consist exactly of the index vectors that belong to the restricted cone. We would ideally like to find a unique representative for each positive orbit and a criterion for an index vector $\mathbf{p}$ to belong to a positive~orbit. 

An elementary argument shows that positive orbits are infinite if $kn>2$: consider the action of $g$ on $\mathbf{p}$. The degree changes as
\begin{align} \label{eq:degree_flop}
\text{deg}(g \mathbf{p}) = (k+1)\; \text{deg}(\mathbf{p}) - (k+2)p_0~.
\end{align}
We can always use the action of permutations to order the components so that $p_0$ is the smallest component, so that $p_0 \leq \text{deg}(\mathbf{p})/(n+1)$. Then we have that
\begin{align} 
\frac{\text{deg}(g \mathbf{p})}{\text{deg}(\mathbf{p})} \geq k+1\;  - \frac{k+2}{n+1}~.
\end{align}
Thus the degree of $g \mathbf{p}$ is larger that that of $\mathbf{p}$ if $kn>2$, showing that if this condition is satisfied, the degree of $\mathbf{p}$ can always be increased, from which it follows that the positive webs must be infinite. In the case $kn<2$, the degree of a positive index vector $\mathbf{p}$ can always be decreased, so there are no positive orbits.

To classify the positive orbits, note that every positive web must have a vector $\mathbf{p}$ of lowest degree. Such a vector satisfies $\text{deg}(g \mathbf{p}) \geq \text{deg}(\mathbf{p})$ for all $g$. Substituting in the expression \eqref{eq:degree_flop} for $\text{deg}(g \mathbf{p})$, it follows that
\begin{align} \label{eq:orderedsource}
p_0 \leq \frac{k}{k+2} \text{deg}(\mathbf{p})~.
\end{align}
This motivates us, after reversing components, to define an \textit{ordered source} as a vector $\mathbf{q}$ that satisfies the above condition and whose components are in decreasing\footnote{We use decreasing order to agree with Ref. \citen{Candelas:2021lkc}. Equation \eqref{eq:orderedsource} only applies to reversed ordered sources.} order $p_{0}\geq \dots \geq p_{n}$. The latter condition is to remove the redundancy introduced by the possibility of permuting the components. Assuming that this `local' condition also implies that the degree of the vector cannot be lowered by any word in $\mathcal{W}$ motivates us to make the following conjecture, which has useful implications for the the structure of the restricted cones, and can potentially be used to study phenomena such as the modularity of prepotentials $\mathcal{F}$ observed in Ref. \citen{Lukas:2022crp}:
\begin{conj} \label{conj:source_conjecture}
Positive orbits are in one-to-one correspondence with ordered source vectors, and each positive orbit contains an ordered source vector as a vector of minimal degree.
\end{conj}
This conjecture is obvious in the two-parameter case. However, with additional parameters, the structure of the Coxeter group is significantly more difficult to keep track of, and we know of no proof in general. However, we have performed extensive computational checks by studying subsets of orbits $\mathcal{W}[\bm p]$ defined by
\begin{align}
\mathcal{W}_{l}[\bm p] \defineas \{w \bm p \; | \; w \text{ a word of length } \leq l \text{ in the generators of } \mathcal{W}\}~.
\end{align}
In every checked case, with various $k$ and $n$, by making $l$ large enough, we have been able to find exactly one source vector in every $\mathcal{W}_l[\bm p]$. For every value of $l$ and every source $\bm s$ we have tested, the set $\mathcal{W}_l[\bm s]$ contains only positive index vectors.
\subsection{A Two-Parameter Example: the Maximal Split of the Quintic}
In the case of two-parameter models that possess an infinite dihedral symmetry, the above conjectures are theorems. For concreteness, we consider the two-parameter CICY given by the maximal split of the quintic threefold\cite{Hosono:2011np,Hosono:2012hc,Hosono:2017hxe}:
\begin{equation}\label{eq:SQ_CICY}
\cicy{\IP^{4}\\\IP^{4}}{1&1&1&1&1\\1&1&1&1&1}~,
\end{equation}
which corresponds to the case $k=4$, $n=2$. In this case the Coxeter group is $\mathcal{W}=\IZ_{2}*\IZ_{2}$, with $*$ denoting the free product of groups. This group has involutive generators $g$ and $s$, and the product $gs$ is of infinite order. Their actions on $\IZ^{2}$ are
\begin{equation}
g:(i,j)\mapsto(4j-i,j)~,\qquad s:(i,j)\mapsto(j,i)~.
\end{equation}
Every element of $\mathcal{W}$ is of the form $s(gs)^{n}$ or   $(gs)^{n}$, $ n\in\IZ_{\geq 0}$. By solving a Fibonacci-like recurrence, the orbit can be written down in closed form:
\begin{align}
(gs)^{n}&\left((i,j)\right)=\notag\\
&\left(i \cosh\left(n \arccosh2\right) + \frac{(2 i - j) \sinh\left(n \arccosh2\right)}{\sqrt{3}}\right.,\notag\\ 
&\hskip20pt\left. j \cosh\left(n \arccosh2\right) + \frac{(i - 2 j) \sinh\left(n \arccosh2\right)}{\sqrt{3}}\right)~.
\end{align} 
The set of positive orbits is $\left\{\mathcal{W}[(i,j)]\right\}_{1\leq i\leq j\leq2i}$. There is a simple algorithm to construct the set of source vectors of positive orbits $\left\{(i,j)\right\}_{1\leq i\leq j\leq2i}$, proving conjecture \ref{conj:source_conjecture} in this case: First note that $\mathcal{W}[(1,1)]$ is a positive orbit. Now we increment the second component by 1 to get (1,2), and $\mathcal{W}[(1,2)]$ is again a positive orbit. Now since $sgs(1,3)=(1,1)$, we have exhausted the set of ordered vectors with first component 1 that represent positive orbits. 

Next we increment the first component, and consider $(2,2)$. We continue to increment the second component to produce more vectors, and stop at $(2,4)$ because $sgs(2,5)=(2,3)$ which we have already tabulated. One continues ad infinitum: increment the second component by 1 until it equals twice the first component, then increment the first component by 1 and repeat. So the sources are $\{(i,j)\}_{0 < i \leq j \leq 2i}$.

Any ordered source vector lies in the cone generated by the two sources $(1,1)$ and $(1,2)$,
since we can write $(i,j)=(2i-j)(1,1)+(j-i)(1,2)$, and $(j-i)$ and $(2i-j)$ are non-negative for $i\leq j\leq2i$. Recall that by the definition of a source, any vector $\bm p$ belonging to the restricted cone can be written as $\bm p = w \bm s$ for a source $s$ and an element $w \in \mathcal{W}$. As any source $\bm s$ can be expressed as a linear combination $\bm s = \lambda_1 (1,1) + \lambda_2 (1,2)$ with $\lambda_i > 0$, any vector $\bm p$ in the restricted cone can be written as $\bm s = \lambda_1 w(1,1) + \lambda_2 w(1,2)$. Therefore we have shown that in this case the restricted cone is generated by the vectors in the orbits $\mathcal{W}[(1,1)]$ and $\mathcal{W}[(1,2)]$. In this case the statement is rather trivial, and many of the vectors in the orbits $\mathcal{W}[(1,1)]$ and $\mathcal{W}[(1,2)]$ do not give independent generators of the restricted cone. However, the same argument can be used to find the restricted cone in terms of sources in cases with more parameters where the structure of the cone is much more intricate, if the conjecture \ref{conj:source_conjecture} holds.
\subsection{A Five-Parameter Example: Mirror Hulek--Verrill Threefolds}
As a much more involved example, consider again the Mirror Hulek--Verrill family \eqref{eq:MHV_CICY}, which corresponds to the case $n=4$, $k=1$.

In this case direct analysis of the orbits of the Coxeter action is difficult due to the number of components of the vectors and the elaborate structure of the corresponding Coxeter group. In particular, finding the set of sources or even the restricted cone has proven to be challenging. However, while we do not have a proof, an extensive computational search suggests that each source can be written as a nonnegative $\IZ$-linear combination of the sixteen sources
\begin{equation}
(1,1,1,0,0)~,\quad(1,1,1,1,0)~,\quad (1,1,1,1,1)~,\quad \text{and permutations.}
\end{equation}
If one assumes the above and conjecture \ref{conj:source_conjecture} then it follows, using the same argument as in the two-parameter case above, that the restricted cone of the family $\MHV$ is generated by the integral vectors belonging to the union 
\begin{equation}\label{eq:genwebs}
\mathcal{W}[(1,1,1,0,0)]\cup\mathcal{W}[(1,1,1,1,0)]\cup\mathcal{W}[(1,1,1,1,1)]~.
\end{equation}
The GV-invariants of degrees $<30$ computed in Ref. \citen{Candelas:2021lkc} are consistent with this, in the sense that, apart from the invariants corresponding to the flopped curves, every non-vanishing invariant belongs to this cone. In this case the converse seems to also be true: every invariant in the cone generated by \eqref{eq:genwebs} is non-zero.

\section{Application: Higher-Genus GV-Invariants}
We have argued that the Coxeter symmetries which arise from symmetries of the manifold and flops to isomorphic manifolds hold for higher-genus invariants as well. Indeed, we can explicitly observe these symmetries in numerous examples for which we have computed the genus-1 invariants, and for the genus-2 numbers computed in Ref. \citen{Hosono:2011np} for the split quintic \eqref{eq:SQ_CICY}. Conversely, the symmetries can be utilized for computing the GV-invariants at higher genera.

The holomorphic anomaly equations (HAE) of Bershadsky, Cecotti, Ooguri, and Vafa \cite{Bershadsky:1993ta,Bershadsky:1993cx} can be used to iteratively compute the genus $g>1$ topological B-model free energy $\mathcal{F}^{(g)}$. The iterative procedure fixes this up to a holomorphic function $f^{(g)}$. This \textit{holomorphic ambiguity} arises because the BCOV recursion relation provides an expression for $\partial_{\overline{\varphi}}\mathcal{F}^{(g)}$, so determining $\mathcal{F}^{(g)}$ up to a holomorphic function of the complex structure moduli $\varphi$. The problem of finding higher-genus GV-invariants is thus reduced to fixing the holomorphic ambiguity.

The A-model free energy $F^{(g)}$, related to the B-model free energy by\footnote{$\varpi_{0}$ is the holomorphic solution in a Frobenius basis about the large complex structure point.}
\begin{equation}
F^{(g)}=\varpi_{0}^{2g-2}\mathcal{F}^{(g)}~,
\end{equation}
can be expanded in terms of the GV-invariants when expressed in terms of the mirror coordinates $t_i$. This expansion is encoded in the formula\cite{Gopakumar:1998ii,Gopakumar:1998jq}
\begin{equation}\label{eq:GVformula}
\sum_{g=0}^{\infty}\lambda^{2g-2}F^{(g)}(\mathbf{t}) = \frac{c(\mathbf{t})}{\lambda^{2}}+l(\mathbf{t})+\sum_{g=0}^{\infty}\sum_{\bm{p}}\sum_{m=1}^{\infty}\frac{n^{(g)}_{\bm{p}}}{m}\left(2\sin\left(\frac{m\lambda}{2}\right)\right)^{2g-2} \ee^{2\pi\ii \bm{p}\cdot\mathbf{t}}~,
\end{equation}
where $c(\mathbf{t})$ and $l(\mathbf{t})$ are the cubic and linear terms that appear at genera 0 and 1. From the free energy $F^{(g)}$, one reads off the genus-$g$ GV-invariants via the relation
\begin{equation} \label{eq:F^g_q-expansion}
F^{(g)}=\sum_{\bm{p}}\left[n^{(g)}_{\bm{p}}+L^{(g)}\left(n^{(h<g)}_{\bm{p}}\right)\right]\text{Li}_{3-2g}\left(q^{\bm{p}}\right)\defineas\sum_{\bm{p}}G_{\bm{p}}\text{Li}_{3-2g}\left(q^{\bm{p}}\right)~, 
\end{equation}
where the second relation defines the quantities $G_{\bm p}$. In this expression, the $L^{(g)}\left(n^{(h<g)}_{\bm{p}}\right)$ denotes a genus-specific linear combination (see e.g. Ref. \citen{Katz:1999xq}) of the lower-genus invariants which appear in the genus-$g$ free energy as bubbling contributions. Since the invariants at any genus are Coxeter symmetric and $L^{(g)}$ is linear, the combination $G_{\bm{p}} \defineas n^{(g)}_{\bm{p}}+L^{(g)}\left(n^{(h<g)}_{\bm{p}}\right)$ also respects this symmetry. 

This invariance of the numbers $G_{\bm{p}}$ can be used to at least partially fix the ambiguity $f^{(g)}$. Namely, we know a priori that $\mathcal{F}^{(g)}$ is a solution to the HAE, such that, when expanded as \eqref{eq:F^g_q-expansion}, the coefficients $G_{\bm p}$ are Coxeter-invariant. Hence, to find the GV-invariants, we can restrict to the set of such solutions. Any two solutions to the HAE differ at most by a rational function $f^{(g)}$, so if we let $\widehat{\mathcal{F}}^{(g)}$ be any solution to the HAE with $q$-expansion given by
\begin{equation}
\varpi_{0}^{2g-2}\mathcal{\widehat{F}}^{(g)}=\sum_{\bm p}\widehat{G}_{\bm p}\text{Li}_{3-2g}\left(q^{\bm p}\right)~,
\end{equation}
with the $\widehat{G}_{\bm{p}}$ Coxeter symmetric, then the remaining holomorphic ambiguity, the difference $f^{(g)} = \mathcal{F}^{(g)}-\widehat{\mathcal{F}}^{(g)}$, also has a Coxeter-symmetric $q$-expansion:
\begin{equation}
\varpi_{0}^{2g-2}f^{(g)}=\sum_{\bm p}\left(G_{\bm p}-\widehat{G}_{\bm p}\right)\text{Li}_{3-2g}\left(q^{\bm p}\right)~.
\end{equation}
This condition constrains the form of the ambiguity significantly. Based on regularity arguments (see for example Ref. \citen{Alim:2012gq}), the ambiguity takes the form
\begin{equation}\label{eq:holamb}
f^{(g)}=\frac{P}{\Delta^{2g-2}Z}~,
\end{equation}
where $\Delta$ is the discriminant locus for the mirror family, and $Z$ is a known polynomial that arises from inverting the Yukawa couplings in the BCOV procedure. Choosing the complex structure moduli suitably, for instance using the choice of coordinates in Ref. \citen{Hosono:1994ax} for toric cases, $P$ is a polynomial in the complex structure moduli of a degree that can be determined by using regularity conditions\cite{Alim:2012gq} of $\mathcal{F}^{(g)}$. One can further restrict the polynomial $P$ by studying the behaviour of the the free energy $\mathcal{F}^{(g)}$ near conifold points in the moduli space. However, generically this is not enough to fully fix the holomorphic ambiguity, and to make progress one has to utilise the known properties of the invariants. For instance, the Castelnuovo bound on invariants, which implies vanishing of invariants of certain degree, has been successfully used to fix the ambiguity to high genera for a class of one-parameter Calabi--Yau manifolds\cite{Huang:2006hq}. Castelnuovo-like bounds \cite{Alexandrov:2023zjb} make it possible to use similar arguments for a wider class of one-parameter manifolds. However, such conditions are not often available, especially in multiparameter cases, so any non-trivial relations between the invariants are invaluable for fixing the ambiguity. 

While the relevance of individual $\IZ_{2}$ identities like \eqref{eq:MHV_inv} was noted in Ref. \citen{Klemm:2004km}, we take a further step by using the infinite group structure given by the existence of multiple such identities. As an example, we consider computation of the genus-2 invariants for the two-parameter split of the quintic corresponding to the CICY matrix \eqref{eq:SQ_CICY}. These have been computed in Ref. \citen{Hosono:2011np}, but we will attempt to avoid using the conifold gap condition, which is technically cumbersome to implement. It turns out that in this case we can set up the problem\footnote{A technical aside: By defining the propagators suitably, the $\IZ_{2}$ symmetry of the problem can be kept manifest throughout so that the holomorphic ambiguity is a symmetric function of the two moduli. The choice of propagators can affect the degree of $Z$ in \eqref{eq:holamb} for some families of manifolds.} so that the polynomial $P$ in the holomorphic ambiguity \eqref{eq:holamb} is a degree-15 symmetric polynomial in the two complex structure moduli. Accordingly, there remain 72 unknowns.

The imposition of Coxeter symmetry fixes the vast majority of the unknown coefficients, leaving only 8 unknowns to fix, corresponding to 8 Coxeter-invariant rational functions of the form \eqref{eq:holamb}. It is possible to fix further constants by using other properties of the GV-invariants. Assuming that the degrees $\bm{p}$ such that ${n^{(1)}_{\bm{p}}=0}$ also have $n^{(2)}_{\bm{p}}=0$, which holds in all known examples, cuts the number of unknowns down to 3. By analyzing such curves on the split quintic, it should be possible to obtain these identities rigorously, as well as the identities $n^{(2)}_{4,4}=0$ and $n^{(2)}_{3,k}=n^{(2)}_{k,3}=0$, which were also obtained in Ref. \citen{Hosono:2012hc} by studying the behaviour of the free energy near singularities. These can then be used to fully fix the holomorphic ambiguity without having to analyze the conifold behaviour of $\mathcal{F}^{(g)}$. Alternatively, the 8 unknown constants left unfixed by Coxeter symmetry should be found by imposing some of the conifold gap conditions.

As another example of constraining the ambiguity with Coxeter symmetries, consider the family
\begin{equation}
\cicy{\IP^{4}\\\IP^{4}}{2&0&1&1&1\\0&2&1&1&1}~.
\end{equation}
In this case we can choose the ambiguity so that the numerator $P$ in \eqref{eq:holamb} is of degree 12, leaving 91 coefficients to fix. Demanding Coxeter symmetry leaves 18~unknowns.

\section{Conclusions and Discussion}
We have demonstrated the existence of a group of symmetries of GV-invariants of symmetric manifolds admitting a flop to a diffeomorphic manifold. In the most interesting cases these symmetries form an infinite Coxeter group, which is of use in computing higher-genus invariants. These symmetries are likely to simplify other problems involving the enumerative geometry of multiparameter Calabi--Yau threefolds, such as black hole microstate counting \cite{Gaiotto:2006wm,Alexandrov:2022pgd,Alexandrov:2023zjb,Gaiotto:2007cd}.

Even though the Coxeter symmetries discussed here explain a significant amount of the structure observed in the GV-invariants of multiparameter Calabi--Yau manifolds, explicit computations of the invariants reveal in many cases additional equalities between the GV-invariants that are not explained by flop transitions to diffeomorphic manifolds. Study of such examples reveals some intriguing patterns that, unlike the Coxeter symmetries discussed here, remain conjectural. Nevertheless, it is tempting to speculate that in several cases there exists a larger symmetry group that the Coxeter group is a subgroup of, and it would be interesting to understand the geometric origin of the conjectural identities.

There are several cases where two distinct Coxeter orbits correspond to the same value of a GV-invariant. In Ref. \citen{Candelas:2021lkc}, it was noted that, being generated by reflections, the action of a Coxeter group on the lattice $\IZ^{h^{1,1}}$ leaves invariant some bilinear form $B$. That is, there exists a bilinear form $B: \IZ^{h^{1,1}} \times \IZ^{h^{1,1}} \to \IZ$ such that
\begin{equation}\label{eq:invform}
B(\mathbf{p},\mathbf{q})=B\left(g(\mathbf{p}),g(\mathbf{q})\right)~, \text{ for all } g \in \mathcal{W}~.
\end{equation}
Such a bilinear form gives a quadratic form $Q(\bm p) \defineas B(\bm p , \bm p)$ that, due to its invariance under the action of $\mathcal{W}$, associates to each orbit a unique integer. An intriguing observation was made for the family of mirror Hulek--Verrill manifolds \eqref{eq:MHV_CICY} in \citen{Candelas:2021lkc}: Let $\bm p$ and $\bm q$ be index vectors that both belong to the restricted Mori cone. If their corresponding GV-invariants agree, $n_{\bm p}^{(g)} = n_{\bm q}^{(g)}$, then their quadratic invariants agree as well, $Q(\bm p) = Q(\bm q)$. This is trivial if $\bm{p}\in\mathcal{W}[\bm{q}]$, but this relation holds even if the vectors $\bm p$ and $\bm q$ belong to distinct orbits of the Coxeter action, and there are many examples of such pairs of distinct orbits. However, the converse is not necessarily true: there are vectors $\bm p$ and $\bm q$ for which $Q(\bm p) = Q(\bm q)$ but $n_{\bm p}^{(g)} \neq n_{\bm q}^{(g)}$. 

This observation makes it tempting to assume that the Coxeter group $\mathcal{W}$ is a subgroup of a larger symmetry group $\mathcal{G}$ which, in turn, is a proper subgroup of the group $O(m,n)$ of transformations that leave the quadratic form $Q$ fixed.

Based on numerous examples of CICYs, analogous relation seems to hold even in the cases where the Coxeter group has rank $r<h^{1,1}$, determining not a unique invariant quadratic form, but instead a $h^{1,1}-r$ parameter family. In these examples, one can uniquely fix the remaining parameters by demanding that if $n_{\bm p}^{(g)} = n_{\bm q}^{(g)}$ for index vectors $\bm p$ and $\bm q$ belonging to the restricted cone, then $Q(\bm p)=Q(\bm q)$. For the examples analyzed, this requirement leads to a system that a priori seems greatly overconstrained. From this perspective it is non-trivial that a solution exists at all.

A particular example of a $Q$-preserving symmetry that does not arise as a part of the Coxeter group is observed for many complete intersection Calabi--Yau manifolds. These are cases where for certain index vectors $\bm p$, the GV-invariants stay invariant under an affine transformation $\bm p \mapsto \bm p + m\bm k$, $m \in \IZ_{\geq 0}$. For instance, for the mirror Hulek--Verrill manifolds, an explicit computation of genus-$0$ and genus-$1$ invariants reveals that, at least up to total degree $30$, the invariants corresponding to 
\begin{align}
\bm p = (0,0,1,1,1)~, \qquad \text{and} \qquad \bm p =(0,0,2,2,2)~,
\end{align}
are invariant under the translation $\bm p \mapsto \bm p + m(0,0,2,2,2)$, $m \in \IZ_{\geq 0}$. Such identities are not unique to this example, but can be observed to hold for various CICYs, for which the invariants can be computed to even higher total degrees. Such symmetries are in fact expected from certain swampland conjectures\cite{Rudelius:2023odg,Lee:2019wij} which relate the invariants to Kaluza-Klein modes appearing at infinite distance limits in the moduli space. Curiously, the Euler characteristic of the mirror threefold always appears as one of the GV-invariants under the translation. This makes it tempting to speculate that there is a mirror symmetry explanation for these symmetries, along the lines of the argument in section \ref{sect:mirror_symmetry}, where the GV-invariant corresponding to the class of curves that are flopped could be deduced from a mirror symmetry argument.

\section*{Acknowledgments}
We thank the organizers of the very interesting ``Gauged Linear Sigma Models @30'' conference in the Simons Center for Geometry and Physics for their work and the opportunity to attend. It is a pleasure to acknowledge numerous conversations with Philip Candelas and Xenia de la Ossa during collaboration on various projects, in particular Ref. \citen{Candelas:2021lkc} which this manuscript is largely based on. We thank Marieke and Peter Tunstall van Beest, and Pietro Ferrero, for hospitality and vehicular assistance on the otherwise insurmountable Long Island. We thank Mohamed Elmi for interesting conversations on the Long and Staten Islands. PK wishes to thank Fabian Ruehle for instructive and illuminating conversations. JM thanks Johanna Knapp and Emanuel Scheidegger for helpful discussion. We thank Hans Jockers for interesting conversations and comments on the manuscript. JM was supported by EPSRC studentship \#2272658, with his research now supported by a University of Melbourne Establishment Grant. PK is supported in part by the Cluster of Excellence Precision Physics, Fundamental Interactions, and Structure of Matter (PRISMA+, EXC 2118/1) within the German Excellence Strategy (Project-ID~390831469).

\bibliographystyle{ws-ijmpa}
\bibliography{proceedings}       

\end{document}